\documentclass[twocolumn,nofootinbib,preprintnumbers,amsmath,amssymb]{revtex4}

\usepackage{graphicx}% Include figure files
\usepackage{dcolumn}% Align table columns on decimal point
\usepackage{bm}% bold math
\usepackage{amsmath}

\def \be {\begin{equation}}
\def \ee {\end{equation}}

\begin{document}

\title{Fermi Pulsars and the Cosmic Ray Knee}% Force line breaks with \\

\author{Andrei Gruzinov}

\affiliation{ CCPP, Physics Department, New York University, 4 Washington Place, New York, NY 10003
}

\begin{abstract}

Nuclei, if lifted into the radiation zone of a weak axisymmetric pulsar, are accelerated to quasi-monoenergetic spectrum, with more than half of the nuclear luminosity emitted nearly monoenergetically, at the energy $E\approx 4Ze\left({L_{\rm sd}\over c}\right)^{1/2}$, where $L_{\rm sd}$ is the spin-down power. 

With a number of (not obviously wrong) assumptions, this turns Fermi (-like) weak pulsars into prominent accelerators at energies near the cosmic ray knee. The main uncertainty is purely theoretical: one needs to understand if nuclei are major positive charge carries in some weak pulsar magnetospheres. 

\end{abstract}

\maketitle

\section{Introduction}

That pulsars can be cosmic ray accelerators is an old idea \cite{Cheng1986} (see also \cite{Bednarek2005} for a relevant discussion of pulsars at the knee and for further references; our study must have been provoked by \cite{Blandford2013}). Now that the weak pulsar magnetosphere has been calculated and the Fermi telescope \cite{Fermi2013} has catalogued the pulsars (and confirmed the theory), this old idea needs an update.

We show in \S\ref{accel} that a weak axisymmetric pulsar of spin-down power $L_{\rm sd}$ accelerates nuclei of charge $Z$ (if these are lifted into the radiation zone)  to energies 
\be
E\approx 4Ze\left({L_{\rm sd}\over c}\right)^{1/2},
\ee
or, in astrophysical notation, 
\be
E\approx 0.7ZL_{34}^{1/2}{\rm PeV}.
\ee

Now consider (a dark counterpart of) one of the brightest pulsars as seen from Earth, PSR J0633+1746. This is a weak nearly-axisymmetric pulsar with spin-down luminosity $L_{34}=3.3$. It therefore accelerates nuclei to about 
\be
E_{\rm Fe}=33{\rm PeV},~~~E_{\rm C}=7.6{\rm PeV},~~~E_{\rm He}=2.5{\rm PeV}.
\ee

If the (dark) pulsar does the nuclear acceleration with the same efficiency with which PSR J0633+1746 accelerates electrons, then the expected cosmic ray flux at Earth is\footnote{ Gamma-rays are beamed, cosmic rays are not; but, as in the standard Cosmic Ray Burst argument, it is also true that invisible pulsars contribute to cosmic ray flux. Incidentally, in the case of PSR J0633+1746, which is close to axisymmetry, the averaged efficiency is actually close to 100\%.}
\be
F=4\times 10^{-9}{{\rm erg}\over {\rm cm}^2\cdot {\rm s}}\times f_{\rm diff}\times f_{\rm sources},
\ee
where $f_{\rm diff}$ and $f_{\rm sources}$ are dimensionless factors, both $\gg 1$. The first factor, $f_{\rm diff}$, accounts for the diffusive (rather than ballistic) propagation of cosmic rays: unlike free-streaming photons, whose energy density scales as $d^{-2}$ with the distance from the source $d$, diffusive propagation of cosmic rays gives the scaling $d^{-1}$ (the diffusive flux through a sphere of radius $d$ does not depend on $d$). The second factor, $f_{\rm sources}$, comes from the same $d^{-1}$ scaling of the energy density of cosmic rays emitted by a source at a distance $d$. As the number of sources scales with the distance as $d^\alpha$ with $\alpha$ between 2 and 3 (probably closer to 2), the distant sources actually dominate. 

\begin{figure}[bth]
  \centering
  \includegraphics[width=0.48\textwidth]{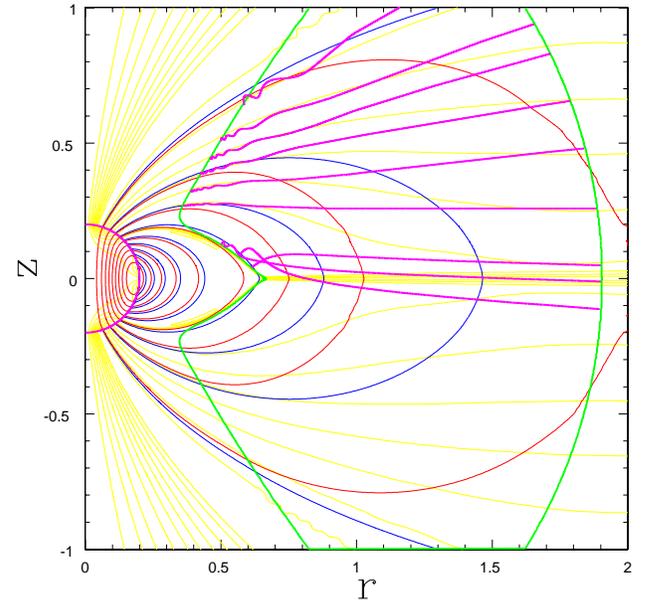}
\caption{Weak axisymmetric pulsar. Thick green -- radiation zone boundary and the artificial outer boundary for propagating the nuclei. Thick magenta -- star and trajectories of the nuclei. Yellow -- poloidal current. Blue -- poloidal magnetic field. Red -- electrostatic potential. } \label{nuclacc}
\end{figure}

Arbitrarily taking $f_{\rm diff}\times f_{\rm sources}=100$, we get the cosmic ray flux $F=4\times 10^{-7}{{\rm erg}\over {\rm cm}^2\cdot {\rm s}}$, which compares interestingly to the observed cosmic ray flux at the knee $\Phi \approx 2\times 10^{-7}{{\rm erg}\over {\rm cm}^2\cdot {\rm s}\cdot {\rm sr }}$.

We actually got $f_{\rm diff}\times f_{\rm sources}\sim 100$, assuming some (not obviously wrong) diffusion model, but reproducing the argument here would be pointless -- diffusion at these high energies (and on such small distances) is obviously highly uncertain. Besides, the true bottleneck is not diffusion, it is a (supposedly) purely theoretical question: {\it do some pulsars use nuclei as major positive charge carries}?

\section{Acceleration}\label{accel}
We took an axisymmetric weak pulsar magnetosphere \cite{Gruzinov2013}, placed nuclei at the boundary of the radiation zone (with small random initial velocities; the initial positions on the radiation zone boundary were chosen so as to get the nuclear current in the radiation zone proportional to the known positron current), and numerically calculated the resulting motion and the final energy. We used the leftover magnetosphere calculated in a small box of size 2x4 (light cylinders); to account for further acceleration we arbitrarily multiplied our final energy (reached at distance 1.9 from the pulsar) by 1.5. The results should be about 50\% accurate. The accuracy can be easily improved, but again this is not the bottleneck.

The final energies for the 9 trajectories shown in the figure were (in units of 
$Ze\left({L_{\rm sd}\over c}\right)^{1/2}$): 3.5, 3.7, 3.7, 2.5, 1.9, 1.4, 1.1, 0.86, 0.58.

\section{Conclusion}
If nuclei are major positive charge carries in some weak pulsars, then Fermi (-like) weak pulsars can well be prominent sources of cosmic rays at the knee energies.

\begin{acknowledgments}

I thank Andrei Beloborodov and Mal Ruderman for telling me that some pulsars can use nuclei as major positive charge carriers. (In the era of ``precision pulsology'', this claim needs to be confirmed by a full calculation of nuclear transport, and showing that it occurs without usual plasma avalanches which could ground the nuclei.) I thank George Pavlov for telling me about neutron star atmospheres (can be anything from He to Fe, and maybe even protons; can be altogether absent). 

\end{acknowledgments}

\end{document}